\documentclass[showpacs,amsmath,amssymb,prl,superscriptaddress,footinbib,twocolumn]{revtex4-1}
\usepackage{amssymb}
\usepackage{graphics}
\usepackage{epstopdf}
\usepackage{epsfig}
\usepackage{dcolumn}
\usepackage{bm}
\renewcommand{\vec}[1]{\mathbf{#1}}
 
\newcommand{\abs}[1]{\left| #1 \right|} 
\newcommand{\avg}[1]{\left< #1 \right>} 
\renewcommand{\d}[2]{\frac{d #1}{d #2}} 
\newcommand{\dd}[2]{\frac{d^2 #1}{d #2^2}} 
\newcommand{\pd}[2]{\frac{\partial #1}{\partial #2}} 
 
\let\baraccent=\= 
\renewcommand{\=}[1]{\stackrel{#1}{=}} 
 
\newcommand{\appropto}{\mathrel{\vcenter{ \offinterlineskip\halign{\hfil$##$\cr\propto\cr\noalign{\kern2pt}\sim\cr\noalign{\kern-2pt}}}}}

\usepackage{color}
 \definecolor{blue}{rgb}{0,0,1} 
 \definecolor{sepia}{rgb}{0,0.8,0.2}
 \definecolor{redi}{rgb}{0.8176,0.0078,0.0078}

\usepackage{ulem}
\begin{document}

\title{ Pinned-to-sliding transition and structural crossovers for helically confined charges}



\author{A. V.  Zampetaki}
\author{J. Stockhofe}
\affiliation{Zentrum f\"{u}r Optische Quantentechnologien, Universit\"{a}t Hamburg, Luruper Chaussee 149, 22761 Hamburg, Germany}
\author{P. Schmelcher}
\affiliation{Zentrum f\"{u}r Optische Quantentechnologien, Universit\"{a}t Hamburg, Luruper Chaussee 149, 22761 Hamburg, Germany}
\affiliation{The Hamburg Centre for Ultrafast Imaging, Luruper Chaussee 149, 22761 Hamburg, Germany}

\date{\today}

\begin{abstract}
We explore the non-equilibrium dissipative dynamics of a system of identical charged particles trapped on a closed helix. 
The particles are subject to an external force accelerating them along the underlying structure.
The effective interactions between the charges induce a coupling of the center-of-mass to the relative motion which in turn
gives rise to  a pinned-to-sliding transition with increasing  magnitude of the external force.
In the sliding regime we observe an Ohmic behaviour signified by a constant mobility. Within the same regime
a structural transition of the helical particle chain takes place with increasing the helix radius leading to a global 
change of the crystalline arrangement. The resulting crystal is characterized by the existence of multiple defects whose number increases
 with the helix radius.

\end{abstract}

\pacs{66.30.-h, 05.70.Ln, 37.10.Ty, 37.90.+j, 45.05.+x}
\maketitle

\begin{center}
 \textbf{I. INTRODUCTION} 
\end{center}
Non-equilibrium is ubiquitous  in nature, since it is rather unlikely for a naturally occurring system to be  completely isolated from its environment.
The continuous interaction of a system with its surroundings brings into play different dissipation mechanisms as well as various external forces driving  it out of equilibrium.
Dissipation and driving are fundamental ingredients for the description and understanding of macroscopic complex effects 
such as friction, lubrication and adhesion constituting the so-called field of tribology \cite{Urbakh2004}. In particular, the majority of the models used in tribological studies
consist of interacting particles subjected to  damping and an external (periodic) potential, and being driven by an external force \cite{Vanossi2013,Cule1996,Baumberger2006,Braun2009}. In the simple case of the well-known
Frenkel-Kontorova (FK) model  \cite{FKbook} the interactions are considered to be harmonic and the driving force constant. Depending on  the height of the external potential,
its commensurability with the interaction bond length and the value of the damping coefficient such systems can exhibit different dynamical behaviors such as an elastic
or a plastic flow, a pinned-to-sliding transition, hysteresis and superlubricity \cite{FKbook, Braun1997, Consoli2000, Tekik2005, Yang2010}.

Relevant physical systems exhibiting such a complex non-equilibrium behaviour include vortices in superconductors \cite{Jensen1988,Koshelev1994,Blatter1994,Reichhardt1996}, colloidal crystals \cite{Reichhardt2004} or Wigner crystals \cite{Reichhardt2001}.
The latter category includes the highly controllable case of trapped ions which can form crystals of different shapes and sizes \cite{Dubin1993,Kjaergaard2003,Drewsen2003a,Drewsen2003b,Knoop2015} resulting from the interplay 
between  the Coulomb repulsion and the external potential. If the latter is periodic, realized for instance by an optical lattice, it has been shown that most of the properties
of the standard FK models can be retrieved \cite{Garcia2007,Puttivarasin2011,Zhirov2011} and the corresponding systems are predicted to exhibit interesting transport properties \cite{Puttivarasin2011,Zhirov2013,Liebchen2015}.
Even more, due to their cleanness and the variety of controllable parameters and structures such systems  offer a  good opportunity to experimentally study 
the numerous effects predicted for  FK models as well as further phenomena connected to microscopic friction or nanofriction \cite{Benassi2011, Mandelli2013,Bylinski2015}.

In the aforementioned cases the interplay between the interactions and an external (periodic) potential is crucial for the systems to exhibit a complex non-equilibrium dynamics.
Vanishing interactions would lead to a lack of collective response whereas a vanishing external potential would in principle allow the particles to move freely 
for arbitrarily small external forces. There exist, however, particular cases where the interactions alone can give rise to a complex dynamics.
These include cases in which the interacting particles (here identical charges) are confined to a one-dimensional curved manifold. If this confining manifold
constitutes a helix, the respective systems of helically confined charges can exhibit many intriguing effects already  in equilibrium.
It is known for example that identical charges in such systems interact through an oscillatory effective two-body potential allowing for the creation of multiple bound states 
 tunable by the geometry parameters \cite{Kibis1992,Schmelcher2011}. In the presence of an external force field perpendicular to the helix axis, superlattice properties arise \cite{Kibis2005a,Kibis2005b} as well as a wealth
of bifurcation phenomena \cite{Plettenberg2016}. If the helix is not homogeneous the interactions additionally couple the center-of-mass (CM) to the relative coordinates \cite{Zampetaki2013},
 which for the specific case of a torus-like closed helix (as studied in the present work) has been found to induce structural phase transitions, deformations of the Wigner crystals' vibrational spectra and highly nonlinear dynamics, all controlled by the geometry \cite{Zampetaki2015a,Zampetaki2015b}.

 The above considerations raise the question: what would be the effect of an external driving force acting on a system of helically confined charges  that exhibit a coupling of the CM 
to the relative coordinates? Are the  constraint-induced deformations of the interactions alone capable of giving rise to effects  associated with static friction such as a pinned-to-sliding transition?
And even more is it possible for such driven systems to exhibit further effects having no analogue in the standard driven FK models?

In order to answer these questions we study here the behaviour of a commensurate system of identical charges confined on a toroidal helix and subject to an external 
force along the helix as well as to damping. We find that due to the coupling of the CM to the relative coordinates there exists a finite barrier 
prohibiting the free motion of the CM. Thus by increasing the force magnitude the system exhibits a pinned-to-sliding transition. The sliding regime is characterized by an Ohmic behaviour,
i.e. the long-time average velocity is proportional to the applied force. The critical force for sliding depends both on the number of particles and on the geometry 
parameters. Tuning the latter and provided that the system is in the sliding regime, a second transition can occur
stemming again from the coupling between the CM and the relative degrees of freedom. This is a rather unexpected transition of a structural type
(precluded in standard FK models)
which results in a global change of the crystalline structure. The emerging structures consist of particle chains of different lengths separated by gaps (defects) 
whose number increases  with the helix radius.

The structure of this work is as follows. We begin in Sec. II by presenting the general equations of motion for the driven-dissipative 
dynamics of $N$ particles confined on a curve and then continue with a description of our specific setup.
Section III presents the features of the steady states reached in the case where the confined particles do not interact.
We return  to the main discussion of the full interacting problem for two and $N$ particles in sections IV and V, respectively.
Finally Sec. VI contains our conclusions and outlook. 

 \begin{center}
 { \textbf{II. SETUP AND POTENTIAL LANDSCAPE}}
\end{center}

Let us consider a system of $N$ identical particles of mass $m$  each, subjected to an external potential $V_F(\vec{r})$ and interacting via a two-body potential 
 $W(\abs{\vec{r}_i-\vec{r}_j})$ that depends only on the interparticle Euclidean distance. Its Lagrangian reads
 \begin{equation}
L(\{\vec{r}_i, \dot{\vec{r}}_i\})=\frac{1}{2} m\sum_{i=1}^N  {\dot{\vec{r}}_i}^2-\sum_{i=1}^N V_F(\vec{r}_i)-\frac{1}{2}\sum_{\substack{i,j=1 \\ i\neq j}}^N  W(\abs{\vec{r}_i-\vec{r}_j}).
 \end{equation}
If the particles are confined onto a smooth, regular  space curve $\vec{r}:\mathbb{R}\mapsto \mathbb{R}^3$
parametrized with the arbitrary parameter $u$, i.e. $\vec{r}_i=\vec{r}(u_i)$, the Lagrangian takes the 
form
\begin{eqnarray}
L(\{u_i, \dot{u}_i\})&=&\frac{1}{2} m\sum_{i=1}^N  \abs{\partial_{u_i}\vec{r}(u_i)}^2\dot{u}_i^2 -\sum_{i=1}^N V_F(\vec{r}(u_i))\nonumber \\
&-& \frac{1}{2}\sum_{\substack{i,j=1 \\ i\neq j}}^N  W(\abs{\vec{r}(u_i)-\vec{r}(u_j)}).\label{lanui}
\end{eqnarray}
 With the Lagrangian in this form it is obvious that the geometry of the constraint manifold affects both the interaction and the kinetic energy, the latter 
 due to the position-dependent factors $\abs{\partial_{u_i}\vec{r}(u_i)}^2$. These factors can be made to become unity \cite{Zampetaki2013}, if desired, by transforming to the arc-length parametrization
$s:u\mapsto s(u)=\int_0^u \abs{\partial_{u'}\vec{r}(u')}du'$. 
The resulting Lagrangian
\begin{eqnarray}
L(\{s_i, \dot{s}_i\})&=&\frac{1}{2} m \sum_{i=1}^N  \dot{s}_i^2 -\sum_{i=1}^N V_F(\vec{r}(s_i))\nonumber \\
&-& \frac{1}{2}\sum_{\substack{i,j=1 \\ i\neq j}}^N  W(\abs{\vec{r}(s_i)-\vec{r}(s_j)}). \label{fueq1}
\end{eqnarray}
consists of kinetic energy terms of the standard Cartesian form at the cost of losing in general 
the explicit analytical form for both the interaction energy and the external potential, as generally the arc-length integral cannot be evaluated in closed form.

The Lagrangian (\ref{fueq1}) yields the equations of motion
\begin{equation}
m \ddot{s}_i=-\pd{V_F(\vec{r}(s_i))}{s_i}-\pd{V(\{\vec{r}(s_i)\})}{s_i}  \label{eomg1}
\end{equation}
where $V(\{\vec{r}(s_i)\})=\frac{1}{2}\sum_{\substack{i,j=1, i\neq j}}^N  W(\abs{\vec{r}(s_i)-\vec{r}(s_j)})$ is the full interaction potential.

Given the familiar Cartesian form of these equations, dissipation can  be added, if desired, in the standard manner 
\begin{equation}
m \ddot{s}_i+ \gamma m \dot{s}_i=-\pd{V_F(\vec{r}(s_i))}{s_i}-\pd{V(\{\vec{r}(s_i)\})}{s_i}, \label{eomg2}
\end{equation}
where $\gamma$ denotes the dissipation coefficient. 

As already mentioned, however, the potential terms in the $s$-representation do not possess in general 
an explicit analytic form, so only an 'approximated version' of eqs. (\ref{eomg2}) can be used in practice.
The equivalent equations in terms of the more convenient $u$ coordinates read
\begin{eqnarray}
m \left(\d{s_i}{u_i}\right)^2\ddot{u}_i&+&  m  \left(\d{s_i}{u_i}\right)\left(\dd{s_i}{u_i}\right)\dot{u}_i^2+ m \gamma \left(\d{s_i}{u_i}\right)^2\dot{u}_i \nonumber \\
&=&-\pd{V_F(\vec{r}(u_i))}{u_i}-\pd{V(\vec{r}(u_i))}{u_i}, \label{eomg3}
\end{eqnarray}
with $\d{s_i}{u_i}=\abs{\partial_{u_i}\vec{r}(u_i)}$. Contrary to eq. (\ref{eomg2}) these equations do always have an analytic expression and they can be integrated
numerically. Nonetheless, the existing nonlinearity in the kinetic energy terms prevents a transparent interpretation and straightforward understanding of the underlying dynamics.
Since the two equivalent sets of equations (\ref{eomg2}) and (\ref{eomg3}) possess complementary advantages, they will be both used in the following to describe
different aspects of the dynamical behaviour of the system under examination.
Let us now proceed to a detailed description of the specific system under consideration.

\begin{figure}[htbp]
\begin{center}
\includegraphics[width=8.4cm]{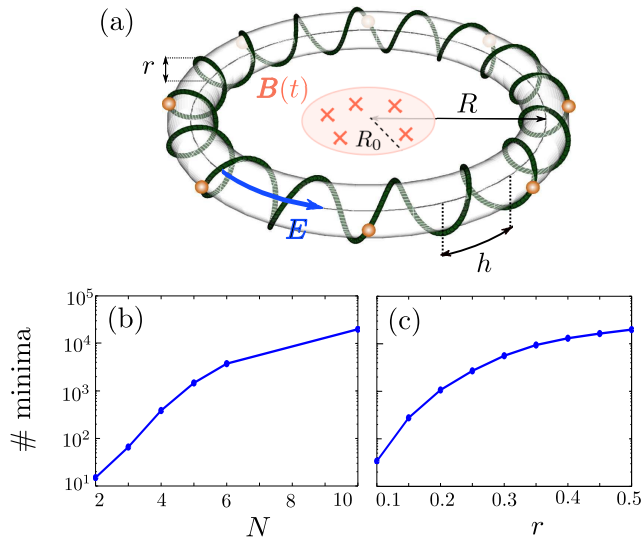}
\end{center}
\caption{\label{tore1} (color online) (a)  Schematic illustration of our setup: A toroidal helix with parameters $R,r,h$ commensurately filled ($n=M/N=2$) with charged particles in their equidistant GS configuration.
A time-varying magnetic field $\vec{B}(t)$ existing in a cylindrical region $R_0<R$ inside the helix induces a circular electric field $\vec{E}$ 
(here assumed to be constant in time) acting on the charges. (b)-(c) An estimation (lower bounds) of the number of stable states (potential minima)  in the field-free case: (b) with increasing number of particles $N$ for
a half-filled ($n=2$) toroidal helix with $r=0.5, h=\pi/2$ and changing $R=N/2$, (c) with increasing radius of the helix $r$ for $N=10, h=\pi/2$ ($R=5$). The results are presented on a semi-logarithmic scale. 
}
\end{figure}

Similarly to our previous works \cite{Zampetaki2015a,Zampetaki2015b}, 
we address a system of $N$ identical charges of mass $m_0$ interacting  via repulsive Coulomb interactions and confined 
to move on a 1D toroidal helix, parametrized as
\begin{equation}
 \vec{r} (u)= \begin{pmatrix}
 \left( R+r \cos u\right)\cos(au) \\
 \left( R+r \cos u\right)\sin(au)\\
 r\sin u
\end{pmatrix}, \quad u \in \mathbb R.
\label{te1}
\end{equation}
 In eq.~(\ref{te1}), $R$ denotes the major radius of the torus (Fig. \ref{tore1}(a)), 
$h$ is the pitch of the  helix and $r$ the radius of the helix. The latter coincides with the minor radius of the torus, such that for $r \rightarrow 0$ the curve reduces to a circle of radius $R$.
These geometric quantities relate to the number of windings $M=\frac{2 \pi R}{h}$ which appears in  eq.~(\ref{te1})
through $a=\frac{1}{M}$.  Correspondingly, $\vec{r}(u)$ is periodic under $u \rightarrow u+2\pi M$.

For such a confining curve the arc-length transformation reads
\begin{equation}
s(u)=\int_0^u \left[a^2(R+r \cos u')^2+r^2\right]^{1/2} du' \label{alp2}
\end{equation}
prohibiting  an analytic closed form expression of the involved potentials in terms of $s_i$.
Still, since the transformation is bijective all the fundamental properties of the effective potentials
on the curved manifold are encoded in their $u$-coordinate representation.

Having the same charge, the particles composing our system are assumed to interact via Coulomb interactions
$ W(\abs{\vec{r}(u_i)-\vec{r}(u_j)})=\frac{\lambda}{\abs{\vec{r}(u_i)-\vec{r}(u_j)}}$ with $\lambda>0$  being the coupling constant. Under the assumption of a helical
confinement the effective two-body potential in terms of $u$ acquires a rather remarkable oscillatory behaviour allowing for the existence of multiple wells
and stable bound configurations which can be tuned in number and strength \cite{Kibis1992,Schmelcher2011,Zampetaki2013, Zampetaki2015a}. 
For more particles the effective potential landscape can have, depending on the geometry parameters of the helical manifold, a high degree of structural complexity \cite{Schmelcher2011}. 

Indications of this high degree of complexity of the interaction potential in our system of charged particles confined on a toroidal helix
can be found in Figs.~\ref{tore1}(b),(c).
In particular, Fig.~\ref{tore1}(b)  presents the number of stable equilibrium states as a function of the number of particles $N$.
Note that  as the number of particles increases the identification of all equilibrium states
becomes a computationally intractable task. Thus, what we observe in Figs.~\ref{tore1}(b),(c) constitutes only a lower bound
for the number of stable equilibria.
Even so, Fig.~\ref{tore1}(b) leads to the conclusion that the number of stable equilibrium states
increases very fast with the number of particles, reaching the order of $10^4$ configurations
already for $N=10$ particles. Complementary information on the potential landscape is presented in Fig.~\ref{tore1}(c), where it is shown that the number of different 
equilibrium configurations increases dramatically also with the increment of the helix radius $r$, i.e. as the toroidal helix moves away from the circle limit. 

The above-mentioned structural complexity of this system provides a strong motivation for examining its non-equilibrium dynamics, e.g. the dynamics under 
the presence of an external force acting on the charged particles. The simplest case is that of an external effective force which is constant in time and the same for all
the particles. A good candidate for this is the electric force $\vec{F}_E$ provided by a time-varying magnetic field 
$\vec{B}(t)=-B(t) \vec{e}_z$ located in a central cylindrical region of radius $R_0<R$, enclosed by the toroidal helix (Fig. \ref{tore1}(a)).
Such a magnetic field with a magnitude increasing in time at a constant rate $\d{B}{t}=\dot B	$ induces
a radially decreasing circular electric field $\vec{E}=\frac{ \dot B R_0^2}{2\rho}  \vec{e}_\phi$, where the use of cylindrical coordinates $(\rho,\phi,z)$ is implied.
This field leads to an external force $\vec{F}_E=\frac{q\dot B R_0^2}{2\rho} \vec{e}_\phi$ acting on each particle possessing a charge $q$. The effective force $F_i$ which the $i$-th
particle feels is finally the projection of $\vec{F}_E$ onto the toroidal helix at the location of the particle, i.e. $F_i=\vec{F}_E \cdot \pd{\vec{r}}{s_i} $. Making use
of $\pd{\vec{r}}{s_i}=\pd{\vec{r}}{u_i} \d{u_i}{s_i}$ and eq. (\ref{te1}) we finally arrive at
\begin{equation}
F_i=\frac{a q \dot B R_0^2}{2} \, \d{u_i}{s_i}=: F \, \d{u_i}{s_i} \label{force1} 
\end {equation}
with $F$ a constant depending among others on the inverse number of windings $a$ and $\d{u_i}{s_i}=\left[a^2(R+r \cos u)^2+r^2\right]^{-1/2}$ (eq.( \ref{alp2})).

Having all the necessary ingredients we can now establish the equations of motion for our specific setup.
We choose dimensionless units by scaling  position $x$, time $t$, energy $E$ and dissipation coefficient $\gamma$ with $\lambda$, $m_0$ and $2h/\pi$  as follows 
 \[\tilde{x}=\frac{x \pi}{2h}, ~\tilde{t}=t\sqrt{\frac{\lambda \pi^3}{8 m_0 h^3}},~ \tilde{E}=\frac{2E h }{\lambda \pi},~\tilde{m}_0=1,~\tilde{\lambda}=1,\]
 \[\tilde{\gamma}=\gamma \sqrt{\frac{8 m_0 h^3}{\lambda \pi^3}}\]
and we omit  in the following the tilde for simplicity. Note that due to this scaling the commonly appearing product $aR=\frac{h}{2\pi}$ is actually fixed to the value $aR=\frac{1}{4}$, but
 still we retain it in the equations for reasons of transparency.

The resulting equations of motion read 
\begin{equation}
\ddot{s}_i+ \gamma \dot{s}_i=-\pd{V(\{\vec{r}(u_i)\})}{u_i}\d{u_i}{s_i}+F \d{u_i}{s_i} \label{eoms1},
\end {equation}
with  $V(\{\vec{r}(u_i)\})=\frac{1}{2}\sum_{\substack{i,j=1, i\neq j}}^N  \abs{\vec{r}(u_i)-\vec{r}(u_j)}^{-1}$, where the use of the inverse transformation $u(s)$ is implied on the r.h.s.

The initial state of the system at $t=0$ is taken to be the ground state configuration with zero particle velocities, corresponding to the absolute minimum of the 
interaction potential $V(\{\vec{r}(u_i)\})$.
For a toroidal helix that is commensurately filled with particles, i.e. $M=nN$, $n=1,2,\ldots$ with $1/n \leq 1$ denoting the filling factor, it is known \cite{Zampetaki2015a}
that for values of the helix radius $r$ up to a critical point $r_c$ the ground state configuration is the equidistant polygonic configuration $u_j^{(0)}=2\left(j-1\right) \pi n$, $j=1,\dots,N$, see Fig.~\ref{tore1}(a). 
 Increasing the helix radius $r$, such a configuration loses its stability at $r_c$ undergoing a zig-zag bifurcation accompanied with an unconventional deformation on the vibrational band structure \cite{Zampetaki2015a}.

 We examine in this work the non-equilibrium dynamics (eq. (\ref{eoms1})) of charged particles confined on a half-filled toroidal helix ($n=2$) in the region $r<r_c$, where the
 initial ground state is still the polygonic one. We assume a very weak dissipation $\gamma=0.01$ and focus our study on $N=60$ particles.
 
The features of the dynamics in this underdamped regime  ($\gamma \ll 1$) are studied
 depending on both the external force magnitude $F$ and on the geometry of the confining manifold controlled by the helix radius
$r$. Before proceeding to a discussion of our results for the interacting many-body system, it is instructive to discuss (sections III, IV)
two simpler cases: 
the non-interacting system and the interacting two-body system.

 \begin{center}
 { \textbf{III. NON-INTERACTING CASE}}
\end{center}
In the case of non-interacting particles ($V=0$) the system of equations (\ref{eoms1}) decouples and reduces to 
\begin{equation}
\ddot{s}_i+ \gamma \dot{s}_i=F \left[a^2(R+r \cos u_i)^2+r^2\right]^{-1/2} \forall i \label{eomni1}
\end {equation}
Since the force term is always positive ($F>0$) the particles are expected to be constantly accelerated. The presence of a small friction 
term should therefore lead to a steady state where the particles move with a constant velocity. 
A more quantitative description requires the expression of eq. (\ref{eomni1}) only in terms of $s_i$.
 For the cases of interest in this work it holds $r \ll R$ since we focus on the regime below the zig-zag transition, characterized by the polygonic force-free ground state described above.
To first order in the small parameter $\frac{r}{R}$ eq.~(\ref{eomni1}) reads as
\begin{equation}
\ddot{s}_i+ \gamma \dot{s}_i=\frac{F}{\sqrt{a^2R^2+r^2}} \left(1- \frac{r}{R}\frac{a^2R^2}{{a^2R^2+r^2}}\cos \frac{s_i}{\sqrt{a^2R^2+r^2}}\right)  \label{eomni2},
\end {equation}
 where we have employed the fact that the arc-length transformation can be written in closed form:
\begin{eqnarray}
s(u)&\approx& \sqrt{a^2R^2+r^2} u+\frac{a^2Rr}{\sqrt{a^2R^2+r^2}} \sin u  \label{appsu1} \\
u(s)&\approx&\frac{s}{\sqrt{a^2R^2+r^2}}-\frac{a^2Rr}{{a^2R^2+r^2}} \sin \frac{s}{\sqrt{a^2R^2+r^2}}. \label{appus1}
\end{eqnarray}
 Since by assumption $ \frac{r}{R} \frac{a^2R^2}{{a^2R^2+r^2}} \ll 1$, to lowest order the oscillating part of the potential in eq.~(\ref{eomni2}) can be neglected leading for long times
to  constant velocity
\begin{equation}
 \dot{s}_i\approx \frac{F}{\gamma \sqrt{a^2R^2+r^2}}. \label{velni}
\end{equation}
Accordingly the $i$-th particle particle coordinate evolves approximately
linearly in time ${s}_i\approx \frac{F t}{\gamma \sqrt{a^2R^2+r^2}}$, yielding within this approximation a constant time averaged force 
$\avg{F_i}= \frac{F}{ \sqrt{a^2R^2+r^2}}$ acting on this particle, see eqs.~(\ref{force1}) and ~(\ref{eomni1}). 

In order to characterize
the moving steady states we define the mobility  $C=\avg{\dot{S}}/\avg{F_S}$, with $S=\frac{1}{N}\sum_{i=1}^N s_i$ being the center of mass 
(CM) coordinate and $F_S=\frac{1}{N}\sum_{i=1}^N {F}_i$ the total force on the CM.  
From the above considerations it turns out that for the cases discussed here we have approximately $C=1/\gamma$, i.e. the 
mobility is independent of both the force and the geometry of the helix, a behaviour known as Ohmic \cite{FKbook}.

\begin{figure}[htbp]
\begin{center}
\includegraphics[width=8.6cm]{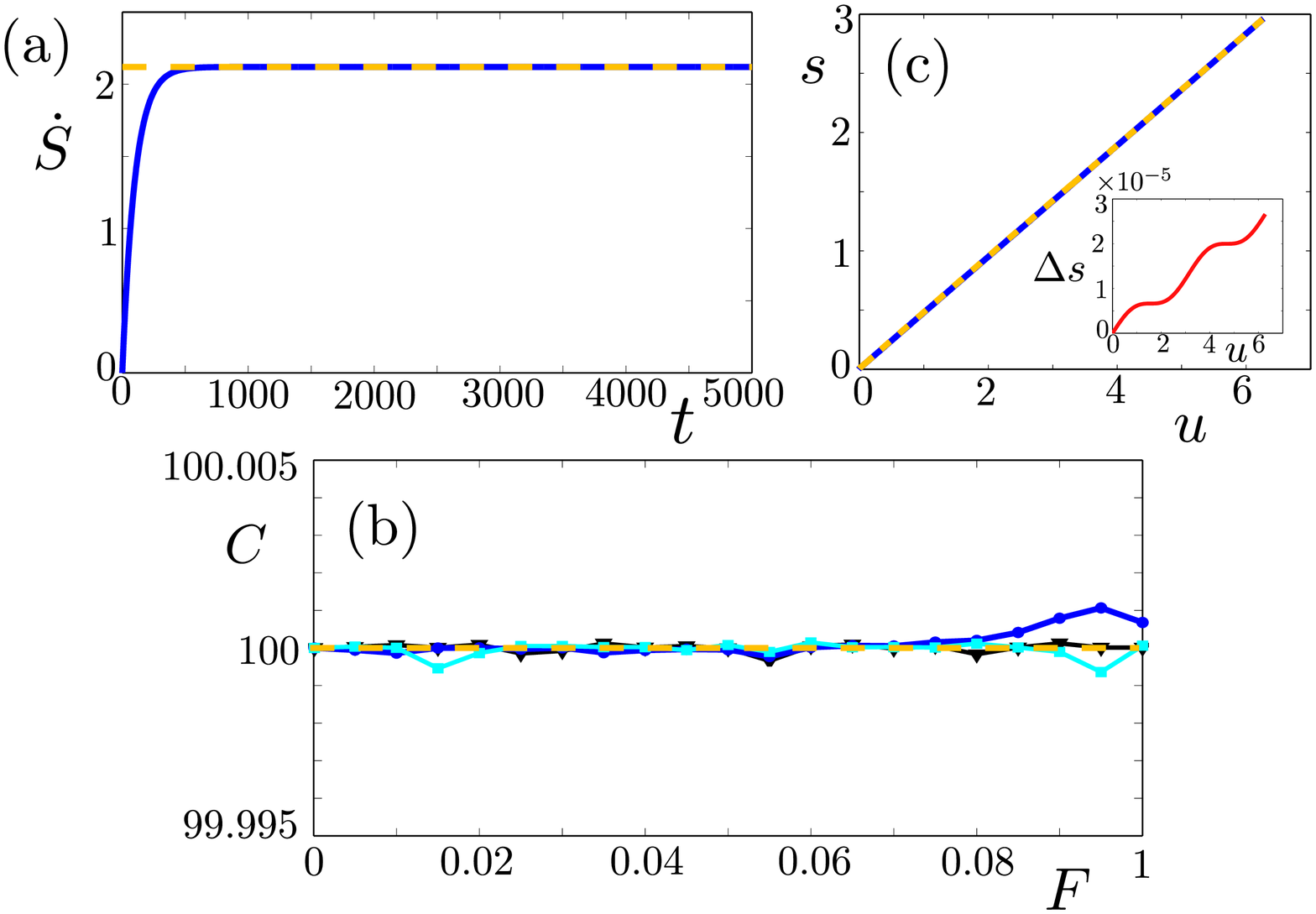}
\end{center}
\caption{\label{nintf1} (color online) (a) The time evolution of the arc-length CM velocity $\dot{S}$ of $N=60$ (thus $R=30$ within our scaling) non-interacting particles  on a toroidal helix with 
$r=0.4$ arranged initially (for $t=0$) in the equidistant polygonic  configuration ($u_j^{(0)}=2\left(j-1\right) \pi n$, $j=1,\dots,N$). The dashed line denotes the analytical prediction for the long-time behaviour expressed in eq. (\ref{velni}). (b) The mobility $C$ as a function of the applied force $F$
for non-interacting systems of $N=60$ particles characterized by different values of the helix radius: $r=0.1$ (black triangles), $r=0.25$ (cyan squares) and $r=0.4$ (blue  circles).
The dashed line marks the analytically predicted value $C=1/\gamma$. (c)  The arc length coordinate $s$ as a function of the $u$-coordinate for $r=0.4$ and $R=30$. 
The dashed line
stands for the approximate expression (\ref{appsu1}) and the inset depicts the difference $\Delta s$ between the approximation and the numerically obtained value.
}
\end{figure}

Numerical results (Fig.~\ref{nintf1}) obtained from the full equations in the $u$-coordinate representation (eq.~(\ref{eomg3})) justify this picture. The CM velocity $\dot{S}$, 
equal here to the velocity $\dot{s}_i$ of each independent particle, saturates at long times to a finite value  which is very well approximated by eq. (\ref{velni}) (Fig.~\ref{nintf1}(a)).
Furthermore, the aforementioned Ohmic behaviour with the mobility $C=1/\gamma$  holds as shown in Fig.~\ref{nintf1}(b) for different values of the helix radius $r$. For reasons
of completeness we present also in Fig.~\ref{nintf1}(c) the behaviour of $s(u)$ in the interval $\left[0, 2\pi\right]$. As expected it is monotonic and for the parameters chosen
(relevant for our discussion in this paper) well approximated by eq.~(\ref{appsu1}).

By including the Coulomb interactions in our model the equations of motion of the individual particles do not only couple to each other but, with the single exception
of the ring ($r=0$), the presence 
of interactions induces a coupling between the CM and the relative degrees of freedom, affecting crucially the particle dynamics. In the following we will investigate the
effect of interactions starting with the simplest case of two confined charged particles.

 \begin{center}
 { \textbf{IV. INTERACTING TWO-BODY SYSTEM}}
\end{center}
We consider a half-filled helix ($M=2N=4$, $R=1$) and  we focus on the charges long-time  dynamics in the underdamped regime
$\gamma=0.01 \ll 1$.

Making use of the explicit form of the external force (eq. (\ref{force1})), the arc-length transformation (eq. (\ref{alp2})), the Coulomb character of the interaction potential
and the parametrization of the toroidal helix (eq. (\ref{te1})) it is straightforward to derive the rather lengthy dimensionless equations of motion both in the $s$ and the $u$ coordinate
representations according to eqs. (\ref{eomg2}),(\ref{eomg3}). In a further step, each set of equations
can easily be combined to give the respective equations of motion for the more convenient
CM $U=(u_1+u_2)/2$ ($S=(s_1+s_2)/2$) and the relative $u=u_2-u_1$ ($s=s_2-s_1$) coordinates.


In the simple
case of a confinement on a ring ($r=0$) the CM decouples from the relative motion and the factors  $\d{s_i}{u_i}$ reduce to constants leading to
\begin{eqnarray}
\ddot{U}+\gamma \dot{U} &=& \frac{F}{a^2 R^2} \nonumber \\
\ddot{u}+\gamma \dot{u} &=& \frac{1}{2^{3/2}aR^3}\frac{\sin a u}{(1-\cos au)^{3/2}}.
\end{eqnarray}
These decoupled equations imply a uniform motion of the CM coordinate and an approach of the relative coordinate to its single fixed point $u=\pi/a$, which
coincides with the equidistant configuration at $t=0$. Therefore Coulomb interacting particles confined on a ring geometry follow the Ohmic
behaviour we have seen in the non-interacting case, i.e.
even the slightest amount of force leads to a sliding of the equidistant crystal characterized at long times by a constant mobility $C=1/\gamma$.

For $r>0$ the CM motion couples to the relative one \cite{Zampetaki2013,Zampetaki2015a}. The first implication of such a coupling 
is the existence of a finite potential barrier originating from the effective interactions,  which the external force needs to overcome in order to set
the CM into motion. It is expected, therefore, that a pinned-to-sliding transition occurs for increasing force magnitude $F$. Such a transition  does  normally not affect 
the interparticle distances. Thus we make the approximation that the relative coordinate remains always close to its equilibrium value $u(0)=4\pi$, i.e. 
$u(t)=4\pi+\delta(t)$ with $\delta(t)\ll 4\pi$,  $\dot{\delta(t)}\ll 1$
making at the same time no additional assumptions on the CM coordinate $U$, i.e we linearize the equations of motion in the relative coordinate. 

Such a linearization leads to a new set of equations where surprisingly the coupling between  the CM  $U$ and the relative  $\delta$ coordinate
is limited only to the equation of motion of the latter, whereas the equation for the CM appears to be completely independent of $\delta$ namely
\begin{equation}
 \alpha(U)\ddot{U}+\frac{1}{2}\alpha'(U)\dot{U}^2+\gamma \alpha(U) \dot{U}=F -\frac{r\sin U}{4\left(R+r \cos U\right)^2}, \label{cmeqU2}
\end{equation}
 with $\alpha(U)=a^2 (R+r\cos U)^2+r^2$ and $\alpha'(U)=\d{\alpha}{U}$. Since the dissipation is by assumption very small $\gamma \ll 1$,
leading to long decay times, the overall
asymptotic behaviour of the system can to a large extent be understood by an investigation of the respective Hamiltonian problem. The effective Hamiltonian for the CM
corresponding to the non-dissipative version of eq.~(\ref{cmeqU2}) reads
\begin{equation}
 H_{CM}(P,U)=\frac{1}{2}\frac{P^2}{\alpha(U)}-F U +\frac{1}{4\left(R+r\cos U \right)},
\end{equation}
where $P$ is the total momentum of the system and the two other terms compose the effective CM potential $V_{CM} (U)$.
This Hamiltonian can explain the behaviour of the CM in terms of a single particle moving in the potential $V_{CM}(U)$.

\begin{figure}[htbp]
\begin{center}
\includegraphics[width=8.6cm]{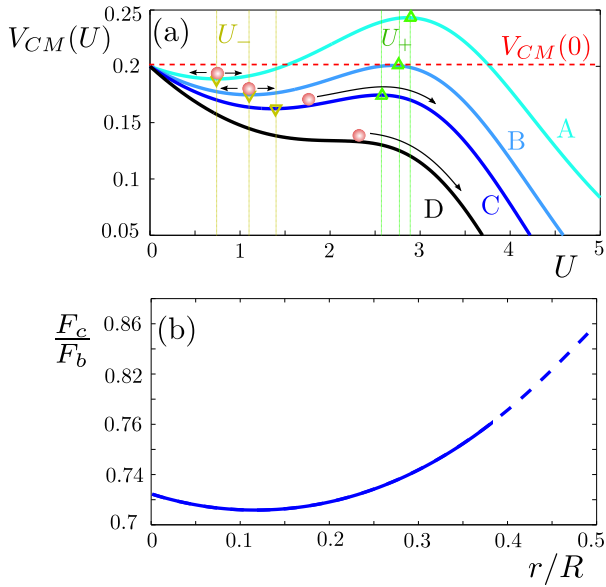}
\end{center}
\caption{\label{cmpot} (color online) (a) The effective CM potential $V_{CM}(U)$ for $r=0.25, R=1$ and for four qualitatively different cases: (A) $F<F_c$, (B) $F=F_c$, (C) $F_c<F<F_b$
and (D) $F>F_b$. The dashed red line marks the initial energy $E=V_{CM}(0)$ whereas the yellow (down triangle) and the green (up triangle) lines mark the positions of minima 
$U_{-}$ and maxima $U_{+}$ respectively. Note that in the case (D) no minimum/maximum exists. (b) The dependence of the critical force $F_c$ (in units of $F_b=\frac{r}{4R^2}$ calculated to lowest order in $\frac{r}{R}$) on the helix radius $r$.
The dashes denote the region where the approximation $r \ll R$ is not valid any more.}
\end{figure}

Given the values of the 
geometry parameters, the dynamics of such a particle can be qualitatively very different 
depending on the magnitude of the external force $F$ (Fig.~\ref{cmpot}(a)). For small values of $F$ (case A)
there exists a large potential barrier which confines a particle starting with zero velocity at the equilibrium point ($U=0$) of the force-free system to the first potential well.
The existence of the small dissipation causes a decay of the  corresponding oscillations, resulting  in the particle relaxation to the minimum $U_{-}$ of the potential for long times. 
In this case we expect therefore a pinning of the particle to the first potential well. At a critical value of the external force $F=F_c$ (case B)
the condition $V_{CM}(0)=V_{CM}(U_+)$ ($U_+$ stands for the maximum of the potential) is satisfied 
and thus the particle stays marginally confined in the potential well. For $F>F_c$ (case C) the value of the potential maximum  
$V_{CM}(U_+)$ becomes lower than the initial energy of the particle $E=V_{CM}(0)$, leading to an escape and an acceleration of the 
particle which due to the dissipation attains finally a finite velocity,  i.e. the sliding regime is entered. 
As $F$ further increases, within this sliding regime, the effective potential  is deformed such that beyond a point $F_b$ the fixed points $U_{+}, U_{-}$ cease to exist (case D). 

The critical value $F_c$ for the pinned-to-sliding transition can be determined given the 
fixed points $U_{+}, U_{-}$, to be derived in the following, by numerically solving the equation  $V_{CM}(0)=V_{CM}(U_+)$ in terms of $F$. The result is shown 
in Fig.~\ref{cmpot}(b) as a function of the helix radius $r$. Note that for reasons of convenience $F_c$ is presented in units of $F_b$, 
containing already an $r$ dependence. Obviously it holds $F_c<F_b$ for every $r$.

Meanwhile the linearized equations for the relative coordinate are far more complex and couple directly to the CM coordinate. Therefore the 
overall behaviour of our system is determined to a large extent by the long-time dynamics of the CM. As we have seen,
this can exhibit two qualitatively different behaviors: pinning or sliding, each prevailing in a 
different regime of external force values $F$. 
In the following we discuss both regimes in more detail, extracting also quantitative information regarding the mobility as well as the
crystalline structure of the resulting steady states.

%
%

 \begin{center}
 \textbf{A. Pinned regime} 
\end{center}
For $F \leq F_c$ the CM is pinned to one of the minima of the effective potential $V_{CM}$, i.e. it relaxes to one of the fixed points of eq.~(\ref{cmeqU2}).
The existence of a fixed point requires
\begin{equation}
g(U)=F-\frac{r \sin U}{4(R+r\cos U)^2}  \label{gU1}
\end{equation}
to become zero. Given that $r \ll R$ for all the cases studied here, we can as a first step keep only the lowest order in $r/R$. This
yields $g(U)=F-\frac{r}{4R^2} \sin U$ which obviously has a root if and only if $F \leq F_b=\frac{r}{4R^2}$. Note that in the underdamped regime  $F_c< F_b$ 
(Fig.~\ref{cmpot}(b)),
meaning that the CM can be sliding even when eq.~(\ref{cmeqU2}) does acquire fixed points.

Returning to the full expression for $g(U)$ (eq.~(\ref{gU1})), the condition for having a root can be rewritten, after squaring, rearranging the terms and introducing $\epsilon = \frac{r}{R}$, $y=\epsilon \cos U$ and $f=\frac{4FR}{\epsilon}$  as
\begin{equation}
 \epsilon^2 f^2(1+y)^4 - \epsilon^2+y^2 =0.
\end{equation}
 The improved value of the critical force $F_b$ can then be found from the condition that the discriminant of this quartic equation vanishes, which yields
 $f_b = 1 + 2 \epsilon^2 + 2 \epsilon^4 +\dots,$
so a slight shift up relative to $F_b =\frac{r}{4R^2}$ is predicted due to  corrections of higher order in $\epsilon$.
However, up to order $\epsilon$ we still have $F_b=\frac{r}{4R^2}$, which we will  use in the following. 
For $F<F_b$, the roots of $g(U)$, corresponding to the potential minima $U_{-}$ and maxima $U_{+}$ shown in Fig.~\ref{cmpot}(a), read to first order in $\epsilon$
\begin{equation}
U_{-} =  \text{arcsin} \frac{F}{F_b} + \frac{2rF}{R F_b}, \quad U_{+} = \pi- \text{arcsin} \frac{F}{F_b} + \frac{2rF}{R F_b}. \label{fixp2b1}
\end{equation}
$U_{-}$, being the stable fixed point of the CM equation (\ref{cmeqU2}), provides an estimation of the steady state into which the system is expected to relax for $F \leq F_c$.

\begin{figure}[htbp]
\begin{center}
\includegraphics[width=8.6cm]{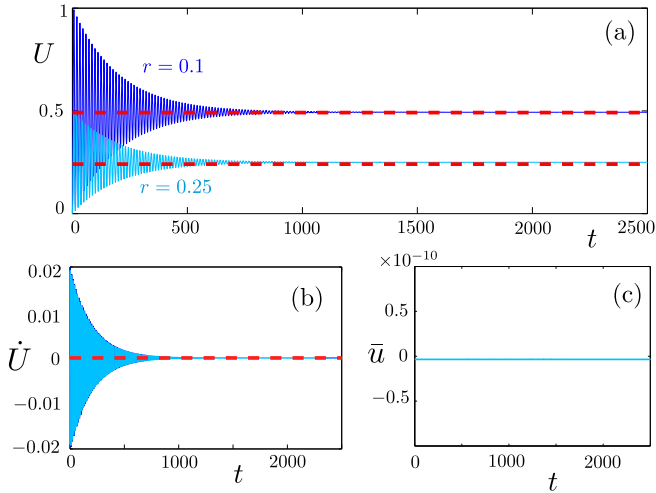}
\end{center}
\caption{\label{pin1} (color online) Time evolution of: (a) the CM coordinate $U$
(b) the CM velocity $\dot{U}$ and (c) the relative coordinate $\bar{u}=u-4 \pi$  for two different values of the radius $r=0.1$ (blue) and $r=0.25$ (light blue), $N=2$  and $F=0.01<F_c$.
The dashed red lines depict the analytical approximations for the steady state $U_-$ (eq.~(\ref{fixp2b1})) to be compared with the initial value of the CM coordinate $U_0=0$.}
\end{figure}

The results of our full numerical simulations shown in Fig.~\ref{pin1} for a specific value of $F \leq F_c$ and two different values of the helix radius $r$ show 
a very good agreement with the analytical
results. The external force $F$ induces oscillations in the CM coordinate $U$ as well as in the CM velocity $\dot{U}$ (Figs.~\ref{pin1}(a),(b)) which decay exponentially in time due to the presence of dissipation.
At long times the velocity of the CM $\dot{U}$ vanishes (Fig.~\ref{pin1}(b)) and the CM coordinate relaxes to a finite value (Fig.~\ref{pin1}(a)), larger than its initial value $U_0=0$, 
which is very well approximated by $U_{-}$ from eq. (\ref{fixp2b1}). 

Inspecting  the behaviour of the relative coordinate $u$, we find that it is almost 
unaffected by the presence of the force (Figs. \ref{pin1} (c)).
This fact can also be explained in terms of the linearized equations of motion. In particular, it turns out that to lowest order
in $\frac{r}{R}$ the  linearized equation for the relative coordinate in the pinned regime ($U \rightarrow U_{-},\dot{U} \rightarrow 0$) reads
\begin{equation}
\ddot{\delta}+\gamma \dot{\delta}=- \frac{rR \cos U_{-}}{4 (a^2R^2+r^2) R^3} \delta.
\end{equation}
Since  $\cos U_{-}>0$  for $F\leq F_c$, 
providing a harmonic restoring force, it follows that the small deviation from equilibrium $\delta(t)$ if not zero initially will be
exponentially decreasing in time relaxing to  $\delta=0$. The equidistant configuration of the particles is therefore always dynamically stable in the pinned phase and the
particles maintain it in the course of their time evolution.
Thus the overall effect of the force in the pinned regime is just to displace the system of particles as a whole, i.e. the displacement of the CM coordinate. 

For $F>F_c$ the particle system possesses sufficient energy to overcome the CM potential barrier induced by the interactions. It enters therefore the sliding regime which
will be discussed and analyzed  in the following.

 \begin{center}
 \textbf{B. Sliding regime} 
\end{center}

In the sliding regime ($F>F_c$) the CM of the particles slides freely around the confining toroidal helix, yielding a uniform motion of the two particles.
The moving steady state of the system can be found by solving the dynamical equation (\ref{cmeqU2}). Due to the nonlinear  prefactors of the derivatives a qualitative understanding
of the long time induced dynamics in terms of eq.~(\ref{cmeqU2}) is hard. We resort therefore
to the equivalent equation for the arc-length CM coordinate $S$.

Making use of the linearization condition $u_i=U+4\pi\pm\delta$ with $\delta\ll 4\pi$, $\dot{\delta}\ll 1$
and keeping overall only the lowest order in $\frac{r}{R}$  (eq. (\ref{appus1})) we derive the equation
\begin{equation}
 \ddot{S}+ \gamma \dot{S}=\frac{F}{\sqrt{a^2R^2+r^2}} \left(1- \frac{F_b}{F}\sin \frac{S}{\sqrt{a^2R^2+r^2}}\right)  \label{eosl1}
\end{equation}
approximating the CM dynamics in the arc-length representation. This kind of equation has been studied extensively in the literature \cite{Coullet2005,Risken1984,Barone1982,Stewart1968,Belykh1977,Liu1981} since it models the behaviour of different systems 
of interest such as a damped pendulum forced with a constant torque \cite{Coullet2005}, non-interacting particles on a periodic lattice \cite{Risken1984} and most importantly the dynamics of Josephson junctions within the so-called Resistively Shunted Junction (RSJ) model \cite{Barone1982,Stewart1968,Belykh1977,Liu1981}. Known features of such systems in the underdamped regime ($\gamma \ll 1$)
studied here are a bistability between equilibrium and periodic solutions for particular values of the parameters and a hysteretic behaviour. 
The bistability occurs in our system in the region $F_c<F<F_b$ where although a fixed point exist, sliding is also possible. Among others this
is expected to lead to a sharp pinned-to-sliding transition at $F=F_c$.

\begin{figure}[htbp]
\begin{center}
\includegraphics[width=8.6cm]{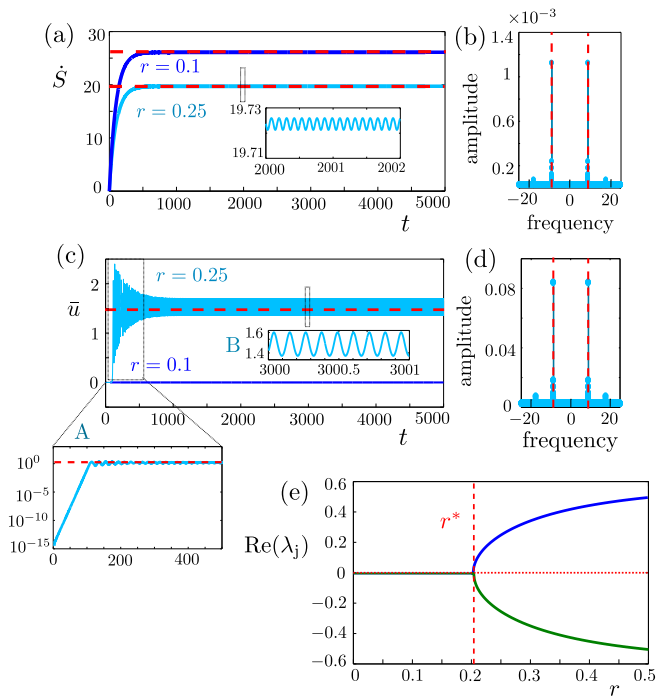}
\end{center}
\caption{\label{slif2} (color online) (a),(c) Time evolution of (a) the arc-length CM  velocity $\dot{S}$ 
and (c) the relative coordinate $\bar{u}=u-4 \pi$ for two different values of the helix radius $r=0.1$ (blue) and $r=0.25$ (light blue), $N=2$ and $F=0.07>F_c$.
The insets in (a) and (c) are zooms of the $r=0.25$ curves for smaller time intervals. In (c) the two insets depict the initial dynamics in a logarithmic scale (inset A)
and the long-lived oscillations in the steady state (inset B).
(b),(d)  The Fourier spectra of the long-lived oscillations in (a),(c) for $r=0.25$. 
In all the cases the dashed red lines depict the various results of analytical approximations for the steady state as discussed in the text. 
(e) The real part of  the eigenvalues of the Jacobian (\ref{jacN2}) relating to the stability of the polygonic configuration. The vertical line indicates $r^*$ whereas
the horizontal one marks the position of zero.}
\end{figure}

For $F>F_c$ and $\gamma \ll 1$ eq.~(\ref{eosl1}) is known \cite{Barone1982} to   possess long-time solutions characterized by
a constant mean velocity, here $\langle\dot{S}\rangle= \frac{F}{\gamma \sqrt{a^2R^2+r^2}}$. 
As shown in Fig.~\ref{slif2}(a) the numerical results for the arc-length CM velocity at different values of the helix radius $r$ agree very well with this estimate. 
In direct analogy with the non-interacting case this fact implies a constant mobility
$C=1/\gamma$, i.e. the system displays an Ohmic behaviour.

Apart from the mean value of the velocity also the frequency $\nu$ of the small oscillations in the steady state
(Fig.~\ref{slif2}(a), inset) can be estimated analytically. 
This should be essentially very close to  the frequency  of the oscillating force term in eq.~(\ref{eosl1}), yielding  $\nu= \frac{F}{2\pi \gamma (a^2R^2+r^2)}$, which 
also complies with our numerical results (Fig.~\ref{slif2}(b)). It is exactly this frequency, originating from the motion of the CM, which appears also 
in the equations of the relative coordinates causing their long time fluctuations (Fig.~\ref{slif2}(c)). This can be clearly seen in Fig.~\ref{slif2}(d) which presents the Fourier spectrum of
the relative coordinate $\bar{u}=u-u^{(0)}=u-4\pi$ for a particular value of the radius. Note that both $\bar{u}, \delta$ are used throughout this manuscript to denote 
the same quantity i.e. the deviation of the relative coordinate $u$ from its initial value $4\pi$. We just use the convention $\delta \ll 4\pi$ for the linearized problem and $\bar{u}$
for the full problem.

 The behaviour of the relative coordinate for two different values of $r$ is depicted in Fig.~\ref{slif2}(c). 
If $r$ is small the relative coordinate does not change from its initial value in the course of the dynamics, meaning that in the steady state the particles move uniformly 
in their equidistant configuration. In contrast, for larger values of $r$ the relative coordinate $u$ jumps exponentially fast (Fig.~\ref{slif2}(c), inset A) to a new average value,
corresponding to  a configuration where the particles are situated closer to each other than in their initial configuration. We therefore observe, within the sliding regime a rather unusual dynamical 
transition in the relative motion depending on the radius of the helix $r$.

\begin{figure}[htbp]
\begin{center}
\includegraphics[width=8.6cm]{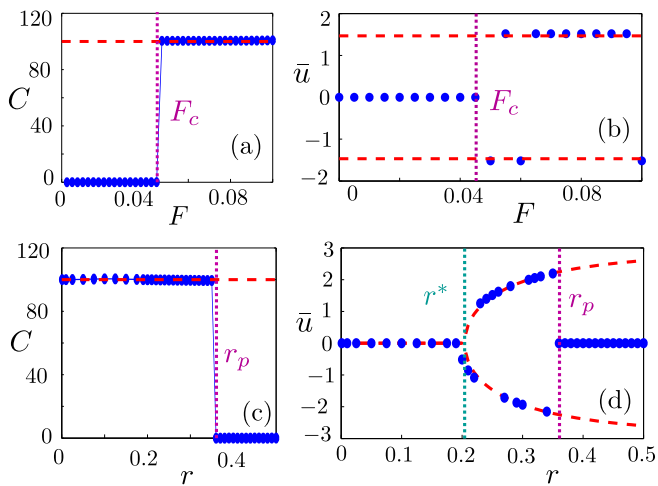}
\end{center}
\caption{\label{fi2all}  (color online) (a) The mobility $C$ as a function of the external force magnitude $F$ for $N=2$ and $r=0.25>r^*$. 
(b) The average value of the relative coordinate $\bar{u}=u-4 \pi$ after the system has reached its steady state as a function of the external force $F$ for $N=2$ and $r=0.25>r^*$.
(c) The mobility $C$ as a function of the helix radius  $r$ for $N=2$ and $F=0.07$. 
(d) The average value of the relative coordinate $\bar{u}$ after the system has reached its steady state as a function of the helix radius $r$ for $N=2$ and $F=0.07$.
The blue dots are the results of our numerical simulations, the red dashed lines the results of our analytical considerations in the sliding regime and the
vertical dotted lines the analytical predictions for $F_c,r_p,r^*$. In all cases  $\gamma=0.01$.}
\end{figure}

In order to unveil the source and the character of this dynamical transition we examine the linearized equation of motion for the relative coordinate $\delta$, the latter being a
rather lengthy equation coupled to the CM coordinate $U$. In the sliding regime the steady state  is characterized by an approximately constant CM velocity $\dot S$,
yielding  in view of eq.~(\ref{appus1}), $U=\frac{F t}{\gamma (a^2R^2+r^2)}$. We remark here that for long times the oscillating terms can be neglected compared to the linear increment. 
As $U$ changes continuously in time at a high rate it is natural to
average this out from the equations of motion for the relative coordinate involving trigonometric functions of $U$. After performing this averaging the linearized equation for the relative coordinate becomes independent 
of $U$ and reads to lowest  order in $\frac{r}{R}$
\begin{equation}
\ddot{\delta}+\gamma \dot{\delta}=- \frac{2a^2 R^2-3r^2}{8 (a^2R^2+r^2) R^3} \delta. \label{relsl1} 
\end{equation}
Obviously beyond a critical value  $r=r^*=\sqrt{\frac{2}{3}}aR=\frac{1}{2\sqrt{6}}$ in our scaling  the elastic constant of the harmonic force term changes sign, altering the character of 
the solutions of eq.~(\ref{relsl1}). The general solution is a linear combination of functions $e^{\lambda_j t}$ where $\{\lambda_j\}$ denote the eigenvalues of the Jacobian $\mathbf{J}$
\begin{equation}
\mathbf{J}= \begin{pmatrix}
  0 & 1  \\
  K & -\gamma 
 \end{pmatrix},~~K=- \frac{2a^2 R^2-3r^2}{8 (a^2R^2+r^2) R^3}.\label{jacN2} 
\end{equation}
of the first-order system of equations equivalent to eq.~(\ref{relsl1}).
The real part of the eigenvalues $\lambda_j$ is shown in Fig.~\ref{slif2}(e) as a function of the helix radius $r$. For $r<r^*$ we have $\rm{Re}(\lambda_1)=\rm{Re}(\lambda_2)<0$
pointing to solutions that decay in time relaxing finally to $\delta=0$. We note that since $\gamma$ is small, both $\rm{Re}(\lambda_j)$ are only slightly smaller than zero here. 
In contrast, for $r>r^*$ the real part of one eigenvalue becomes positive yielding solutions
that increase exponentially in time, in accordance with the  numerically observed initial time evolution of the relative coordinate $\bar{u}$ shown in Fig.~\ref{slif2}(c).
In order to find the final value of the relative coordinate characterizing the steady state for $r>r^*$ we need to  relax the assumption that $\bar u$ $(= \delta)$ is small.
With $u= 4 \pi+\bar{u}$, $a \bar{u} \ll 1$ (the angular inter-particle separation is bounded above by the inter-winding separation) and $r \ll R$
the effective equation for the relative coordinate after averaging out the CM $U$  reads
\begin{equation}
\ddot{\bar{u}}+\gamma \dot{\bar{u}}=- \frac{1}{8 (a^2R^2+r^2) R^3}(2a^2R^2 \bar{u}- 3r^2 \sin\bar{u}) \label{relsl2} 
\end{equation}
which obviously reduces to eq.~(\ref{relsl1}) for $\bar{u}=\delta \ll 1$. The fixed points of this equation,  $\sin \bar{u}=\frac{2a^2R^2}{3 r^2} \bar{u}$, provide the
possible steady-state configurations. For $r<r^*$ the only fixed point is $\bar{u}=0$. This becomes unstable at $r=r^*$ where the system undergoes a pitchfork bifurcation leading to two 
stable non-zero solutions  of opposite sign implying a symmetry breaking. These solutions turn out to approximate well the actual mean values of the relative coordinates in the steady state  for $r>r^*,~F>F_c$ (Fig.~\ref{slif2}(c)).
Having analyzed the dynamics of the interacting two-body system it is instructive to briefly summarize our understanding before we move on
to the interacting many-body system on the toroidal helix.
 \begin{center}
 \textbf{C. Conclusions for the two-body system} 
\end{center}
Figure \ref{fi2all} illustrates our results for the underdamped dissipative dynamics of the two-body system.

The Coulomb interactions in this system couple the CM and relative coordinates. However, by
performing a linearization in the relative coordinate the CM equation of motion decouples from the relative one (eqs.~(\ref{cmeqU2}),(\ref{eosl1})),
indicating that the CM moves approximately independently and that its motion determines to a large extent the overall dynamics of the system. 
For a fixed value of the helix radius $r$, the CM behaves differently depending on the value of the external force $F$. 
If $F<F_c$ (with $F_c$ a critical value) the CM is pinned, the mobility $C$ of the system is zero (Fig.~\ref{fi2all}(a)) and the particles
relax finally due to dissipation into a static equidistant configuration with their CM displaced from its initial value. 

On the other hand if $F>F_c$ the CM potential barrier can be overcome and the CM slides freely around the toroidal helix. In this sliding regime the mobility is constant, $C=1/\gamma$ (Fig.~\ref{fi2all}(a)),
signifying an Ohmic behaviour. The estimate of $F_c$ extracted from Fig.~\ref{cmpot}(b) agrees very well with our  numerical results for the unapproximated equations of motion. 
The transition from the pinned to the sliding regime is very sharp in our case of underdamped dissipative dynamics owing to the existence of the bistability region $F_c<F<F_b$.
 It turns out that $F_c$ depends monotonically on $r$, thus a sliding-to-pinned transition, 
from mobility $C=1/\gamma$ to $C=0$, can also occur for a fixed value of $F$ by increasing $r$ (Fig.~\ref{fi2all}(c)).
Indeed beyond a certain point $r_p$ such that $F_c(r_p)=F$ the system enters the pinned regime.

Surprisingly enough, when the system is still in the sliding regime another transition occurs as the helix radius $r$ is increased, this time affecting the relative coordinate $u$ (or $\bar{u}=u-u^{(0)}$)
(Fig.~\ref{fi2all}(d)). This transition, occurring at the critical value $r^*$ is of a symmetry-breaking pitchfork type and yields for $r>r^*$ two opposite solutions 
 characterized by a reduced interparticle distance. As shown, the analytical estimations both for the critical radius $r^*$ and for the average relative distance in the 
final steady state (eq.~(\ref{relsl2})) are in very good agreement with the numerical results. The small deviation of the latter from the analytics can be attributed to 
higher orders in $\frac{r}{R}$ which become more important close to $r^*$. As expected for $r>r_p$ the system enters the pinned regime where the deviation from the equidistant configuration
is  again zero. Apart from that, as long as the condition $F>F_c$ is satisfied, the average relative distance in the steady state is independent of the value of the external force $F$ 
(Fig.~\ref{fi2all}(b)).

We can understand the mechanism of the above-summarized transition as follows. The external force affects primarily only the CM of the particles and causes them
to run around the toroidal helix at a constant speed without changing their interparticle distance. This motion of the particle system however affects due to the coupling 
also the relative coordinate and it does so in a very particular way.
It affects the average potential the particles feel, so that for long times the particles actually see a new effective potential
which is obtained by averaging in time the CM coordinate. This new dynamical effective potential can have, depending on the geometry parameters, different minima and maxima than
the static one, trapping eventually (with the help of dissipation) the particles in one of them.

Note that  Fig. \ref{fi2all} shows results of independent simulations and no continuous variation of $F$ or $r$ in time has been assumed. By making the force 
$F$ time dependent this system is expected to present a hysteretic behaviour, i.e. for adiabatically increasing $F$ the results of Fig.~\ref{fi2all}(a) are expected to  be observed whereas 
for decreasing $F$ from a large value the sliding is expected to persist even for $F<F_c$ as in the underdamped case of the RSJ model for Josephson junctions \cite{Barone1982} or the underdamped driven FK model \cite{Braun1997,FKbook}. A detailed
analysis of hysteresis in our system goes, nonetheless, beyond the scope of the present work.
 \begin{center}
 \textbf{V. INTERACTING MANY-BODY SYSTEM} 
\end{center}
 
Our focus is on  the long-time behaviour of the $N$-particle system in the presence of a circular electric field  
 (Fig. \ref{tore1} (a)). Although this is far more complex than the two-body case  discussed in detail above, the main features of the dynamical behaviour and the 
 computational steps used to derive the analytical results are surprisingly similar. We will discuss them therefore for completeness
 rather briefly, focusing on the novel dynamical features.
 
 The complexity of the many-body problem manifests itself  already from the very first step of introducing the CM $U=\sum_{i=1}^N \frac{u_i}{N}$ and the $N-1$ relative $\tilde{u}_i=u_{i+1}-u_i$ coordinates. 
 Using these
  \begin{equation}
u_i=U+\frac{1}{N}\sum_{k=1}^{N-1}k\tilde{u}_k-\sum_{k=i}^{N-1} \tilde{u}_k= U+\tilde{U}_i,
 \end{equation}
 where we have introduced $\tilde{U}_i$ standing for the total relative coordinate part of $u_i$. 
 A linearization in the relative coordinates $\tilde{u}_i$ amounts to  
 $\tilde{u}_i= 4\pi+\delta_i(t)$ or $\tilde{U}_i=(2i-N-1)2\pi+\Delta_i(t)$  
 ($\delta_i, \Delta_i \ll 4 \pi,~\dot{\delta}_i, \dot{\Delta}_i \ll 1$).
 
 Using these in the CM equation  we obtain
 \begin{equation}
  \alpha(U)\ddot{U}+\frac{1}{2}\alpha'(U)\dot{U}^2+\gamma \alpha(U) \dot{U}= F -\frac{c_N r \sin U}{4\left(R+r \cos U\right)^2} , \label{cmeqUN}
\end{equation}
 with $\alpha(U)=a^2 (R+r\cos U)^2+r^2$ as introduced above and $c_N=\sum_{l=1}^{N-1} (\sin \frac{\pi l}{N})^{-1}$ being a constant that depends on the number of particles $N$.

Surprisingly, apart from the numerical prefactors in the potential this equation 
is essentially the same as eq.~(\ref{cmeqU2}), and  since $c_2=1$ it reduces exactly to it for $N=2$.
Thus eq.~(\ref{cmeqUN}) implies the same dynamics as the one described in the previous section. In particular there is a critical value $F_c$, 
given  by Fig.~\ref{cmpot}(b) (where the use of the appropriate value of $F_b$ (see below) is implied), at which the system exhibits a pinned-to-sliding transition. Following the structure of the previous section we
discuss first the main characteristics of the dynamics in the pinned and the sliding regime and provide then  a summary of the general dynamical picture.

 \begin{center}
 \textbf{A. Pinned and sliding regime} 
\end{center}

For $F\leq F_c$ the CM of the particles $U$ is pinned in the relevant stable fixed point of eq.~(\ref{cmeqUN}). As in the two-body case
such fixed points exist for $F<F_b=\frac{c_N r}{ 4R^2}$  (to lowest order in $r/R$) with $F_c<F_b$, with a region of bistability  ($F_c<F<F_b$) in which  the particles can slide although
 a stable fixed point is still available. The fixed points of the $N$-body system to lowest order in $\frac{r}{R}$ are given by the expression (\ref{fixp2b1})
 if the  general $N$-dependent value $F_b=\frac{c_N r}{4R^2}$ is used. As in the two-body case,
$U_{-}$ corresponds to the stable fixed point where the system relaxes to for long times,
whereas $U_{+}$ denotes the unstable fixed point. 

\begin{figure}[htbp]
\begin{center}
\includegraphics[width=8.6cm]{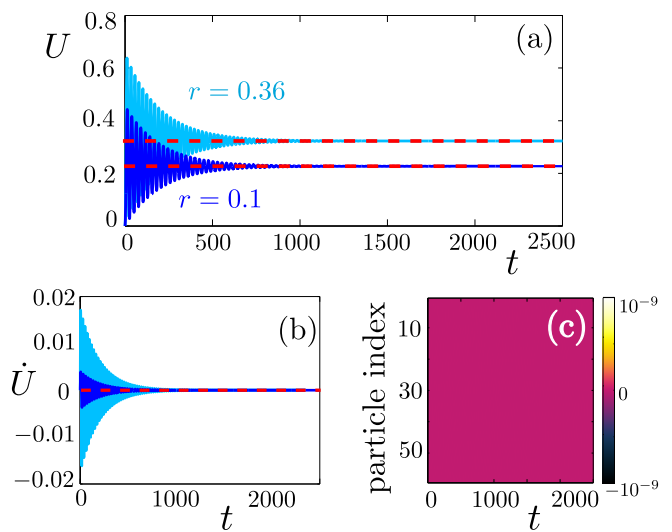}
\end{center}
\caption{\label{pinN60} (color online) Time evolution of: (a) the CM coordinate $U$
(b) the CM velocity $\dot{U}$ for two different values of radius $r=0.1,~F=0.001$ (blue) and $r=0.36,~F=0.005$ (light blue) both for $N=60$.
(c) Time evolution of the relative coordinates $\bar{u}_i=\tilde{u}_i-4 \pi$ (their values are encoded by color) for $N=60,F=0.005, r=0.36$. Obviously 
the relative coordinates do not change in time. Note that  a similar result is obtained for the case $r=0.1$.
The dashed red lines depict the analytical approximations for the steady state (eq.~(\ref{fixp2b1})).}
\end{figure}

 Fig.~\ref{pinN60} demonstrates that these expressions approximate very well the solutions of the full numerical problem for $N=60$ particles. 
The behaviour of the $60$-particle system in the pinned regime matches  the behaviour of the two-particle one: The CM is overall displaced 
from its initial equilibrium by a positive amount (Fig.~\ref{pinN60}(a)), the CM velocity
decays to zero (Fig.~\ref{pinN60}(b)) and the relative distances do not change from their initial values (Fig.~\ref{pinN60}(c)), i.e. the particles remain equidistantly spaced
in a polygonic configuration.
 
\begin{figure}[htbp]
\begin{center}
\includegraphics[width=8.6cm]{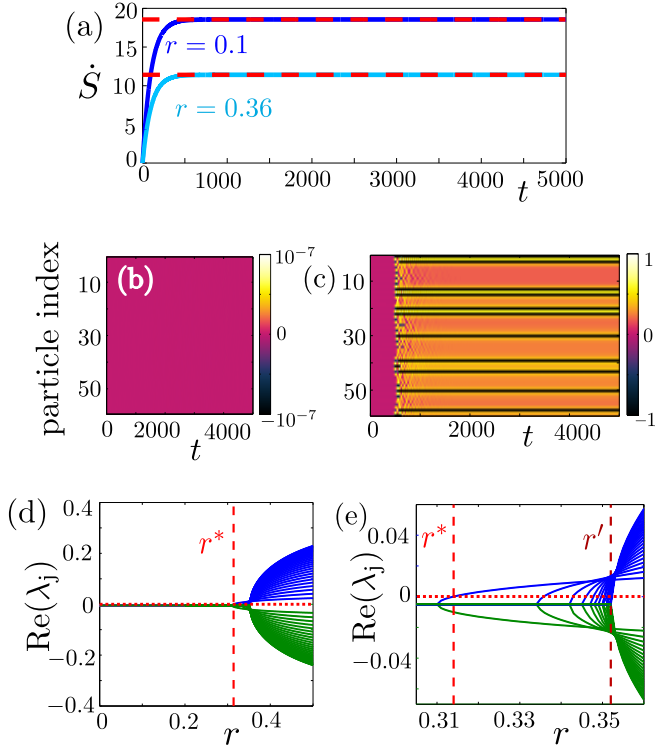}
\end{center}
\caption{\label{sliN60} (color online) (a) Time evolution of the arc-length CM velocity $\dot{S}$ for two different values of radius $r=0.1$ (blue) and $r=0.36$ (light blue), $N=60$  and $F=0.05>F_c$.
 Dashed red lines depict the analytical approximations for the steady state as discussed in the text.
(b),(c)  Time evolution of the relative coordinates $\bar{u}_i$ (encoded by color) for $N=60, F=0.05$ and (b) $r=0.1$, (c) $r=0.36$.
(d) The real part of  the eigenvalues of the Jacobian (\ref{jacNN}) relating to the stability of the polygonic configuration.
The vertical line indicates $r^*$ whereas the horizontal one marks the position of zero.
(e) Zoom of figure (d) for $r$ in the region of $r^*$. The existence of a point $r'$ beyond which $N-1$ of the eigenvalues become positive is obvious.
}
\end{figure}

Regarding the sliding regime the equation of motion in terms of the arc-length coordinate $S$ is again identical to eq.~(\ref{eosl1})
(with the corrected value for $F_b$), meaning that the $N$-body system displays, concerning its CM dynamics, the same features as its two-body analogue. In particular, the 
system relaxes to a moving state in which the particles move at a constant CM velocity $\dot{S}$ with a mean value very well approximated by 
$\avg{\dot{S}}= \frac{F}{\gamma \sqrt{a^2R^2+r^2}}$, see Fig.~\ref{sliN60}(a). Among others this  indicates a constant mobility $C=1/\gamma$.

Concerning the dynamics of the relative coordinates, for small values of the helix radius $r$ they remain unchanged pointing to the crystalline structure keeping its 
polygonic form in the course of time evolution (Fig.~\ref{sliN60}(b)). In contrast, for large enough values of $r$ the relative coordinates change substantially over a finite time
(Fig.~\ref{sliN60}(c)), relaxing thereafter to different values marking the existence of a new crystalline structure to be discussed in more detail below 
(see also Fig.~\ref{helcon}(c)). This is reminiscent of the  abrupt change in the value of the relative coordinate in the two-body case (Fig.~\ref{slif2}(c)), but due to the large number of
particles and therefore the large number of qualitatively different configurations, it is much more complex. From the point of view of the many-body system this abrupt change
in the relative coordinates can be interpreted as a dynamical structural transition induced by the uniform motion of the particles, established by the simultaneous action
of the external force $F$ and the dissipation.
 
The linearized equations accounting for this transition can be obtained in analogy to the two-body problem, i.e. by assuming $\tilde{u}_i= 4\pi+\delta_i(t)$,  
($\delta_i  \ll 4 \pi, \dot{\delta}_i \ll 1$), $r \ll R$ and by averaging out the CM coordinate $U=\frac{F t}{\gamma (a^2R^2+r^2)}$. The  result reads
 \begin{eqnarray}
  \ddot{\delta}_i&+&\gamma \dot{\delta}_i= \sum_{j=1}^{N-1} K_{ij} \delta_j, \nonumber \\
  K_{ij}&=& \begin{cases} f(i-j)-f(i), &  i\neq j \\ -2f(i)-\sum_{l \neq i}^{N-1}f(i-l) & i=j \end{cases} \label{linrelN1}
   \end{eqnarray}
 where
\begin{equation}
f(n) = \frac{a^2R^2\left[3+\cos (4\pi a n)\right]-r^2\left[\frac{5}{2}-\frac{1}{2} \cos (4\pi an)\right]}{4 \sqrt{2}(a^2R^2+r^2)R^3 \left[1-\cos (4\pi an)\right]^{3/2}} \label{funrelN1}. 
\end{equation}
It can be checked that eq.~(\ref{linrelN1}) reduces to eq.~(\ref{relsl1}) for $N=2$. For the general $N$-body case eq.~(\ref{linrelN1}) constitutes
an $N-1$ linear system of second order coupled equations. As known this can be reduced to  a linear system of $2(N-1)$ first-order equations admitting solutions which  evolve in time according to the
exponentials $e^{\lambda_j t}$, with $\lambda_j$ being the eigenvalues of the Jacobian $\mathbf{J}$,
\begin{equation}
\mathbf{J}= \begin{pmatrix}
  \mathbf{O} & \mathbf{I}  \\
  \mathbf{K} & -\gamma \mathbf{I}
 \end{pmatrix},\label{jacNN} 
\end{equation}
where $\mathbf{O}$ denotes the $(N-1)\times(N-1)$ zero matrix and $\mathbf{I}$ the $(N-1)\times(N-1)$ identity matrix. The real parts of the eigenvalues of this matrix computed 
numerically for $N=60$ particles are presented in Fig.~\ref{sliN60}(d). As shown for values of the helix radius $r$ below a critical value $r^*$
we have $\textrm{Re}(\lambda_j)<0~ \forall j$, meaning that the equidistant polygonic configuration (around which we have linearized our equations of motion)
is dynamically stable. For  $r>r^*$ on the other hand at least one eigenvalue has a positive real part, yielding an exponential increase in time and dynamical instability in the direction of the corresponding 
eigenvector. The number of dynamically unstable eigenvectors increases as the radius of the helix $r$ is increased (Fig.~\ref{sliN60}(e)) until it reaches a point $r'$
for which exactly $N-1$ eigenvalues (the same number as the number of relative coordinates) have positive real parts.

The new crystalline structure which emerges and gets stabilized in time when the equidistant configuration is unstable, can in principle be deduced by the generalized
equations for the relative coordinates analogous to eq.~(\ref{relsl2}). These are calculated by averaging out the CM coordinate in the $N-1$ equations of the relative coordinates 
without assuming that the deviation from the equidistant structure is small i.e. without linearizing.

The resulting system of equations is rather complex and due to the existence
of trigonometric terms and the large number of degrees of freedom it supports multiple fixed points for a large number of particles $N$.
The problem of their identification is therefore computationally extremely demanding. For these reasons we restrict our analytical considerations here to 
identifying the regimes of stability 
of the polygonic configuration. We resort in the following
to our numerical results in order to give a phenomenological description of the specific crystalline structure of the resulting steady states
while summarizing the main features of the dynamics of the $N$-body system.

\begin{figure}[htbp]
\begin{center}
\includegraphics[width=8.6cm]{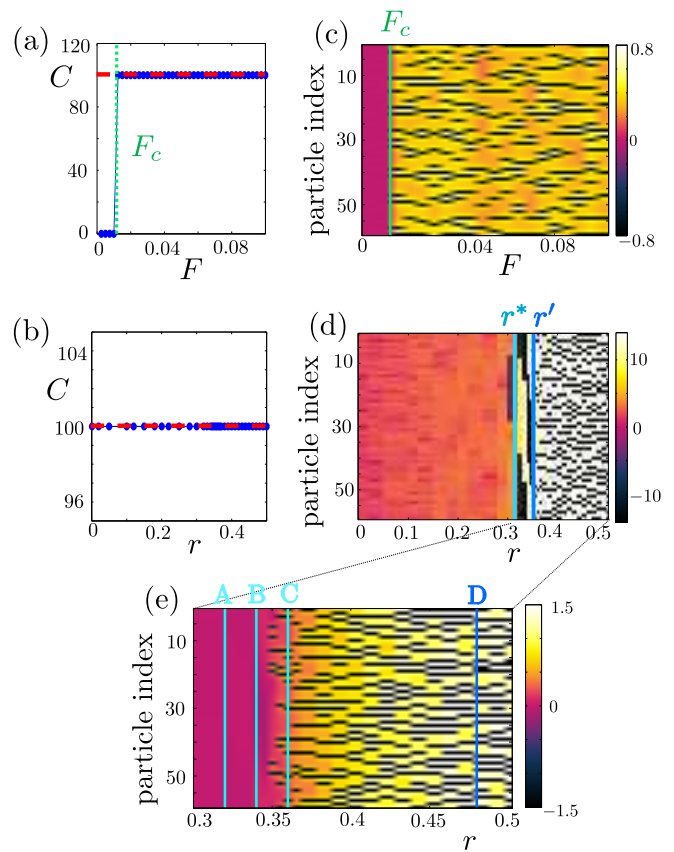}
\end{center}
\caption{\label{fi60all} (color online) (a) The mobility $C$ as a function of the external force magnitude $F$ for $N=60$ and $r=0.36>r^*$. 
(b) The mobility $C$ as a function of the helix radius  $r$ for $N=60$ and $F=0.05>F_c$.
 The analytical prediction $C\approx 1/\gamma=100$ in the sliding regime is indicated by the red dashed lines.
(c) The steady state time-averaged value of the relative coordinates $\bar{u}_i$ (encoded by color) as a function of the external force $F$ for $N=60$ and $r=0.36$.
(d) The steady state  time-averaged value of the relative coordinates $\bar{u}_i$ (encoded by color) as a function of the helix radius $r$ for $N=60$ and $F=0.05$.
The quantity $y_i=\textrm{sgn}\bar u_i(\log |\bar u_i|+14)$ is shown in order to facilitate the identification of the different regimes. 
Note that therefore the values in $[-5,5]$ correspond to $\bar{u}_i \leq 10^{-9} \approx 0$. Magnification: (e) the values of $\bar{u}_i$ are shown for $0.3<r<0.5$.
 Vertical lines in panels (a), (c) and (d) indicate the (semi-)analytical predictions for $F_c$, $r^*$, $r'$.
}
\end{figure}

 \begin{center}
 \textbf{B. Structural transition} 
\end{center}

Figure~\ref{fi60all} summarizes our results for the behaviour of a system of $N=60$ charged particles. 

\begin{figure*}[htbp]
\begin{center}
\includegraphics[width=17cm]{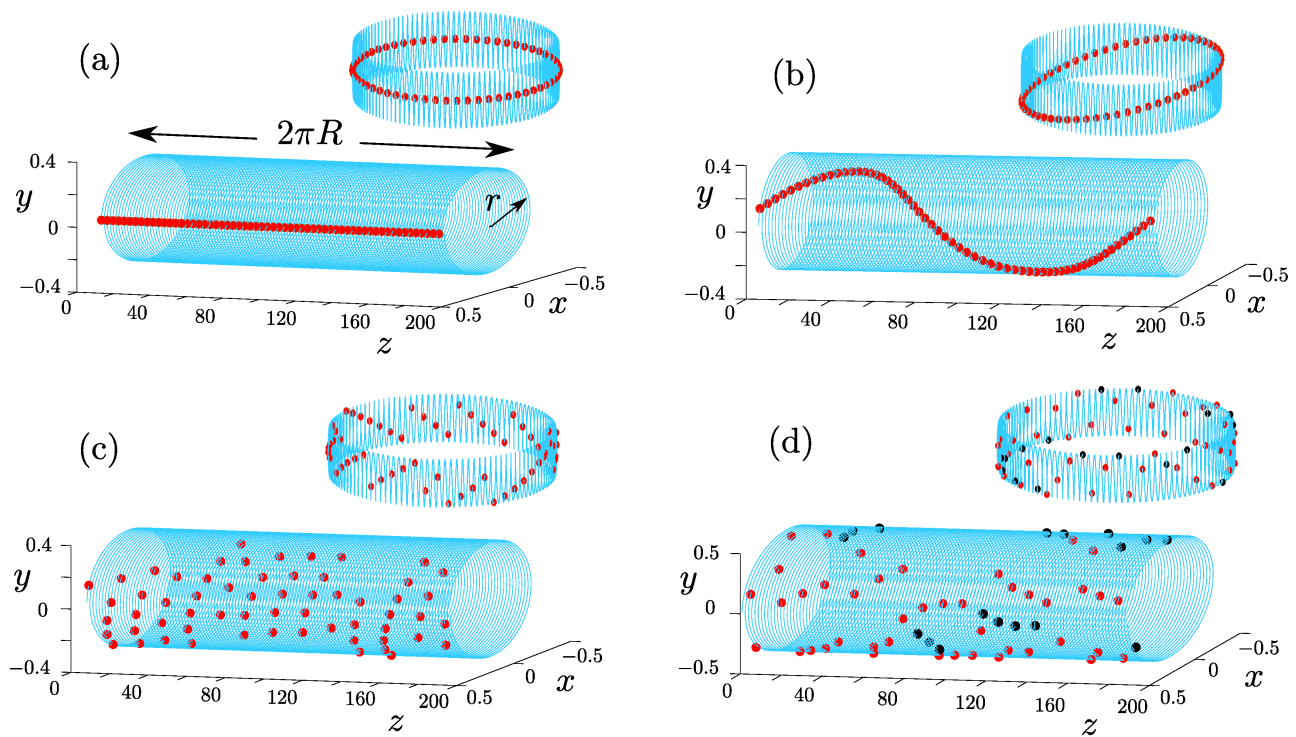}
\end{center}
\caption{\label{helcon} (color online) (a)-(d)  The steady state averaged particle configurations with their CM shifted to $U=U_0=0$ in the sliding regime for $N=60, F=0.05>F_c$
and for increasing helix radius corresponding to the points (A,B,C,D) marked in Fig.~\ref{fi60all}(e). Apart from the configurations on the toroidal helix 
shown in the insets, we present also the same configurations in the equivalent straight helix (obtained by cutting the toroidal helix at a point and making it straight).
This rather peculiar representation is chosen to allow the distinction between the inner and the outer side of the helix, i.e. 
to make clear the contribution of the helix radius $r$ which is overshadowed in the representation of the toroidal helix due to $r \ll R$.
Red points stand for particles occupying  for $U=0$ the outer region of the helix whereas black for particles in the inner region.
}
\end{figure*}

As in the two-body case, for a fixed value of the radius $r$ the mobility $C$ jumps at a point $F_c$ from zero value, signifying the pinned phase,
to a non-zero value $C \approx 1/\gamma$, independent of the driving force, characterizing the sliding phase (Fig.~\ref{fi60all}(a)).
In this phase all particles and consequently their CM travel around the helix at the same almost constant average velocity for long times. The latter is proportional to the average external force,
a so-called Ohmic behaviour.
The predicted value for the critical force $F_c$ as deduced by the effective Hamiltonian for the CM coordinate (Fig.~\ref{cmpot}(b)) agrees very well with the results of the 
full simulations. Note that also in the $N$-body case the pinned-to-sliding transition is very sharp owing to the existing bistability in the region $F_c<F<F_b$.
The possibility for sliding already for $F>F_c$ overshadows the smooth transition expected for $F>F_b$, shifting the critical point to a lower value and imposing
a discontinuity in the pinned-to-sliding transition.

Concerning the dependence of the mobility $C$ on the helix radius $r$, when the system is in the sliding regime, the mobility 
appears to remain approximately constant at the value $1/\gamma$ (Fig.~\ref{fi60all}(b)) matching the analytical 
predictions, i.e. the mobility is independent of the helix radius. 

The most interesting part of the system dynamics is the dynamical structural transition occurring in the sliding phase for $r>r^*$ (Figs.~\ref{fi60all}(c),(d)). Here
the many-body system can display much richer characteristics than that of the two particles owing to the huge number of possibilities for a structural change.
For $F>F_c$ and a fixed radius $r>r^*$ we can see that the magnitude of the external force essentially has no effect on the  qualitative character (e.g. average number of defects)
of the steady state reached through the 
structural transition (Fig.~\ref{fi60all}(c)).  Given this overall qualitative similarity, there is still a plethora of quantitatively different possible final states  (e.g. different locations of defects) that may result from the structural transition.
Which one is finally realized strongly depends on the exact initial conditions leading to the behaviour observed in  Fig.~\ref{fi60all}(c). This is to be compared 
to the two-body case where the only possible states resulting from the symmetry breaking are two (Fig.~\ref{fi2all}(b)) with opposite values as in the standard case 
of a pitchfork bifurcation. Although the dynamically stabilized states in the many-body case are essentially very difficult to be described and predicted,
they seem to share some common characteristics: in the relative coordinate picture they consist of an approximately constant number of localized defects on top of a more or less uniform background  
(Fig.~\ref{fi60all}(c)).

The emergence of the structural transition for a fixed value of the force magnitude as a function of the helix radius $r$ is presented in Fig.~\ref{fi60all}(d), where 
for clarity a logarithmic scale has been used. As shown beyond a critical value $r^*$, well approximated by the result of the stability analysis in the previous paragraph,
the crystalline structure of the steady state changes substantially from the initial equidistant configuration presented in Fig~\ref{helcon}(a). In the first stage of 
the transition and within a small regime $r^*<r<r'$ the structural change amounts to a smooth  modulation of the relative coordinates (Fig.~\ref{fi60all}(d)).
A typical crystalline configuration at this stage is presented in Fig.~\ref{helcon}(b). This is obtained by time-averaging the particles position once in the steady state 
and shifting their CM to $U=0$. Obviously, the configuration has  a smooth shape reminiscent of a second harmonic mode 
with all the particles occupying for a given value of the CM  only one part of the toroidal helix (for $U=0$ the outer part).
The occurrence of such a structure is  plausible from the near-linear stability analysis above. 
For $r^*<r<r'$ only a small number of the Jacobian eigenvalues have positive real parts (Fig.~\ref{sliN60}(e)), resulting in only a few unstable modes
which  yield smooth modulations in the final structural shape. 

For $r>r'$ the situation is very different. Now the periodic structure is unstable in all directions and the resulting steady states possess a rather complex structure 
characterized as we have already mentioned by a number of defects on an approximately uniform background of interparticle $u$-distances (Fig.~\ref{fi60all}(d)).
An inspection of these steady state configurations in the $3D$ space (Fig.~\ref{helcon}(c),(d)) 
leads us to the conclusion that these defects are in fact gaps between particle chains of different lengths. For smaller values of $r$, these chains have similar lengths and lie
on one part of the toroidal helix (for $U=0$ the outer part) forming an overall ordered structure (Fig.~\ref{helcon}(c)). In contrast, for larger $r$ values the resulting crystal 
has a rather disordered structure and some particles occupy  for $U=0$ also the inner part of the toroidal helix (Fig.~\ref{helcon}(d)).

\begin{figure}[htbp]
\begin{center}
\includegraphics[width=7.6cm]{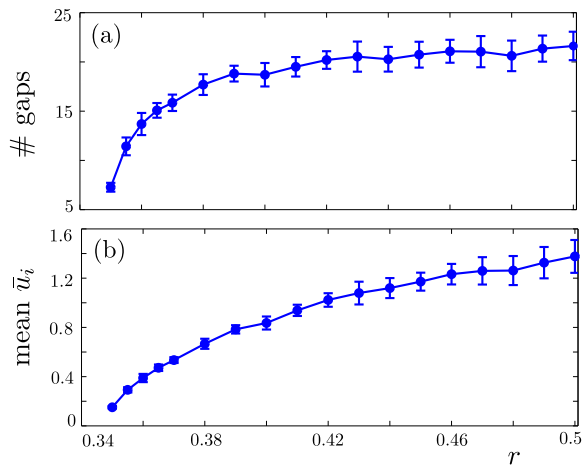}
\end{center}
\caption{\label{gaps1}  (color online) (a) The mean number of gaps and
(b) the mean interparticle distance $\bar{u}_i$, ignoring the gaps, as a function of the helix radius for $N=60,F=0.05>F_c$ in the regime $r>r'$.
All the results plotted are obtained by averaging over $30$ independent realizations. The error bars included stand for $\pm \sigma$ deviation from the mean values, with $\sigma$ denoting the standard deviation of the ensemble of the different realizations.}
\end{figure}

Returning to Fig.~\ref{fi60all}(d), it is evident that the principal feature in the regime $r>r'$ is the increment of the number of gaps (or defects) with the increment of the radius $r$, see Figs.~\ref{fi60all}(d),(e). 
This increment (after statistically averaging over different realizations) turns to be rather smooth and characterized by a saturation at large values of $r$, see Fig.~\ref{gaps1}(a).
Meanwhile the mean interparticle $u_i$ distance in the crystalline structure (ignoring the gaps),
is also seen to increase smoothly (Fig.~\ref{gaps1}(b)), compensating for the increment of the number of gaps.
 In both Figs.~\ref{gaps1} (a),(b) the statistical errors are small enough to justify our conclusions.

\begin{figure}[htbp]
\begin{center}
\includegraphics[width=6cm]{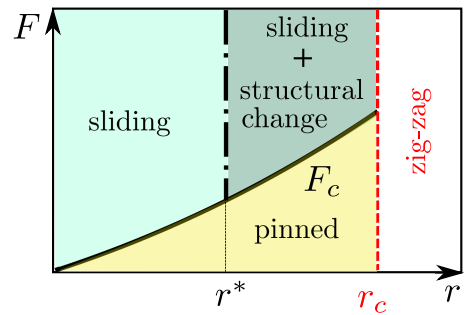}
\end{center}
\caption{\label{conc1} (color online) A schematic illustration of  the ``phase diagram'' of our finite system of $N$ charges confined on a toroidal helix
of radius $r$ under the presence of an external force of magnitude $F$.}
\end{figure}

 \begin{center}
 \textbf{VI. CONCLUSIONS AND OUTLOOK}  
\end{center}
We have demonstrated that a system of charged particles confined on a toroidal helix exhibits an intriguing dynamical behaviour when an 
external force along the helix acts on all particles  in the presence of a small dissipation.  In typical similar setups, such a force would just accelerate the CM of the system.
Here, however, the effective interactions among the charges cause a coupling of the CM to the relative motion leading to a more diverse dynamical 
behaviour, depending both on the value of the external force magnitude $F$ and on the value of the helix radius $r$. This behaviour is,
as  summarized in Fig.~\ref{conc1}, mainly characterized by two qualitatively different transitions.

The first transition concerns the long-time dynamics of the CM as the force magnitude $F$ is increased. For small values of $F$ the CM potential 
barrier stemming from the effective interactions cannot be overcome by the force and due to the presence of dissipation the CM is finally pinned to
one of the  effective potential wells, i.e. the crystal is just displaced globally by a finite amount keeping its initial structure. 
For $F$ beyond a critical value $F_c$, depending on the geometry parameters of the system, the external force has sufficient magnitude in order to 
overcome the potential barrier and set the CM into motion (sliding regime in Fig.~\ref{conc1}). 
Due to the small dissipation the CM attains finally a finite CM velocity proportional
to the acting force, i.e. the system exhibits an Ohmic behaviour.  Such a  behaviour of the CM constitutes a typical example of the so-called pinned-to-sliding transition.

The second transition relates to the final structure of the sliding crystal. 
The uniform motion of the CM  in the sliding regime changes the average effective potential that the charged particles feel. 
For values of radius $r$ greater than a critical value $r^*$ the initial equidistant configuration of the crystal becomes unstable
and a new state, corresponding to a minimum of
the new, time averaged effective potential, becomes dynamically stabilized. 
With increasing $r$, therefore, we encounter a structural transition  within the sliding regime (Fig.~(\ref{conc1})). 
For a two-body system such a symmetry-breaking transition is signified by a pitchfork bifurcation, resulting in the two particles coming closer to each other for $r>r^*$.
For a many-body system this is far more complex, since the new effective potential landscape possesses multiple structurally different minima. Numerical evidence shows that
the states that are finally realized consist of a number of particle chains of different lengths separated by large gaps. Interestingly the average number of
these gaps is found to increase with the helix radius $r$.
Note that despite the structural change, the average CM velocity of the crystal is intact.

This second kind of a dynamical structural transition occurring on a sliding crystal is rather unusual since it requires 
an explicit coupling of the CM to the relative coordinates, which is rarely found
in the models commonly used to examine mechanical out-of-equilibrium situations, such as those of the Frenkel-Kontorova type \cite{FKbook}.
In our system the coupling between the CM and the relative motion is not only present
but it also allows for a  clear separation of time scales, leading to a CM motion that is on average independent of the crystal structure
and to a structure that is determined by the time-averaged potential induced by the CM motion.

Regarding the possibility of an experimental realization of our setup, it primarily relies on the challenging task of constructing a helical trap for charged particles.
Although such a construction might not be straightforward, there have been certain advances towards this direction in different contexts, such as the realization
of helical nanostructures \cite{Prinz2000,Zhang2009,Lee2014,Ren2014} and the proposal of optical helical traps for neutral atoms \cite{Bhattacharya2007,Reitz2012,Okulov2012}.
Moreover, macroscopic realizations of our system could also be possible by using for example charged
beads as done for the study of polymers \cite{Reches2009,Tricard2012}. 
Depending on the final size of the system, a further challenge would be to provide
a time-varying magnetic field of high enough rate of change in order to be able to observe the discussed effects of the pinned-to-sliding transition. 
Nevertheless, let us note at this point that the results presented here are rather robust with respect to the exact type of the external force used, 
as long as this is approximately constant along the helical manifold. Therefore, it  is an open question whether different particularly more easily realizable setups 
which exhibit 
equivalently rich dynamics do exist.

From the theory side, further studies could be dedicated to the analysis of the system's behaviour in different parameter regimes, e.g.
in the overdamped case and in the regions close to the transition points ($F_c, r^*$). These are expected to give further insight into the character
of the two observed transitions, especially if complemented with the study of the expected hysteretic behaviour in such a system.
Along these lines, an exploration of the dynamics of charged particles on a closed helix under the effect of a time varying force is an intriguing problem. 


\begin{center}
{ \textbf{ACKNOWLEDGEMENTS}}
\end{center}
A. V. Z. thanks G. M. Koutentakis and P. G. Kevrekidis for fruitful discussions.

\end{document}